\documentclass[12pt,epsfig]{article}

\input epsf
\usepackage{amssymb}
\usepackage[dvips]{graphicx}
\usepackage{amscd}
\usepackage{mathrsfs}

\def\dj{\hbox{d\kern-0.347em \vrule width 0.3em height 1.252ex depth
-1.21ex \kern 0.051em}}

\begin{document}

\setlength{\oddsidemargin}{0cm}
\setlength{\baselineskip}{7mm}


\thispagestyle{empty}
\setcounter{page}{0}

\begin{flushright}
CERN-TH/2003-100   \\
HU-EP-03/19\\
{\tt hep-th/0305093}
\end{flushright}

\vspace*{2cm}

\begin{center}
{\bf \Large General properties of noncommutative field theories}

\vspace*{1cm}

Luis \'Alvarez-Gaum\'e \footnote{E-mail: {\tt Luis.Alvarez-Gaume@cern.ch}}
and Miguel A. V\'azquez-Mozo
\footnote{E-mail:
{\tt Miguel.Vazquez-Mozo@cern.ch}}$^{,}$\footnote{On leave from
F\'{\i}sica Te\'orica, Universidad de Salamanca,
Salamanca, Spain.}

\vspace*{0.5cm}

{\sl Theory Division, CERN \\
CH-1211 Geneva 23 \\
Switzerland}

\vspace*{1cm}
{\bf \large Abstract}
\end{center}

\noindent
In this paper we study general properties of noncommutative field theories obtained
from the Seiberg-Witten limit of string theories in the presence of an external
$B$-field.  We analyze the extension of the Wightman axioms to this context
and explore their consequences, in particular we present a proof of the CPT theorem
for theories with space-space noncommutativity.  We analyze as well questions 
associated to the spin-statistics connections, and show that noncommutative
$\mathcal{N}=4$, U(1) gauge theory can be softly broken to $\mathcal{N}=0$ 
satisfying the axioms and providing
an example where the Wilsonian low energy effective action can be constructed
without UV/IR problems, after a judicious choice of soft breaking parameters
is made.  We also assess the phenomenological prospects of such a theory,
which are in fact rather negative.

\newpage

\section{Introduction and motivation}
\setcounter{equation}{0}

Noncommutative geometry has had a profound influence in Mathematics  
(cf. \cite{math}). Also in Physics it has
had an effect in a number of subjects ranging from applications 
to the integer and fractional quantum Hall effects \cite{hall,Ezawa} in  
condensed matter physics \cite{condensed}, to
the noncommutative formulation of the standard model by Connes and  
Lott \cite{ConnesLott} (see also \cite{revCL} for
a review).

In string theory the use of noncommutative geometry was  
pioneered by Witten \cite{WittenOS}
in his formulation of open string field theory. Compactifications of  
string and M-theory on noncommutative
tori were studied in \cite{CDSDH}. Shortly after this, Seiberg and  
Witten \cite{SW} realized that a certain class
of quantum field theories on noncommutative Minkowski space-times can  
be obtained as a particular low energy
limit of open strings in the presence of a constant NS-NS $B_{\mu\nu}$  
field (see also \cite{b-field}).
This result generated a flurry of activity in the study of quantum  
field theories in noncommutative
spaces (see \cite{revNCG} for reviews). Part of the interest has been  
aimed at getting new insights
into the regularization and renormalization of quantum field theories  
in this novel framework.
In fact, many features of ordinary (commutative) field theories have  
found rich analogues in the noncommutative context
\cite{solitons}.

The noncommutative field theories obtained from string theory via the  
Seiberg-Witten limit are
neither local nor Lorentz invariant, since the fields in the action  
appear multiplied with Moyal products.
Locality, together with Lorentz invariance, have been traditionally considered  
to be two of the holy principles 
in a quantum field theory. With few exceptions,  
there has been little motivation to try to
extend the principles of quantum field theory to non-local (or  
non-Lorentz invariant) theories.
Although in general allowing for non-locality in quantum field theory  
creates havoc, the Seiberg-Witten
limit yields a very specific class of non-local theories where,  
together with some unusual features,
some of the desirable properties of local theories are preserved.  
Therefore it has been a challenge to
understand the consequences, both theoretical and phenomenological, of  
these theories \cite{Wg}.

The first unexpected property of noncommutative field theories was  
pointed out by Minwalla,
van Raamsdonk and Seiberg \cite{MvRS}. These authors realized  
that quantum theories on noncommutative spaces
are afflicted from an endemic mixing of ultraviolet and infrared divergences.  
Even in
massive theories the existence of ultraviolet divergences induce  
infrared problems \cite{IR/UV}. As a consequence the Wilsonian approach to field theory seems to break  
down: integrating out high-energy
degrees of freedom produces unexpected low-energy divergences,  
inducing in the infrared operators
of negative dimension. This lack of decoupling of high energy modes  
seem to doom these theories from
any phenomenological perspective (see Section~\ref{phenomenology}).

In ordinary Quantum Field Theory there are important consequences that  
follow from the general principles
of relativistic invariance and locality, which do not necessarily  
extend to the noncommutative case.
Results like the CPT theorem \cite{cpt,Jost:CPT,Streater:vi} and the  
spin-statistic connection
\cite{ss} will not necessarily hold. Similarly, questions concerning  
the existence of an $S$-matrix,
its unitarity and the notion of asymptotic completeness are in need of  
drastic revisions.
In Refs. \cite{SJ,Carroll:2001ws,ajsw,Chaichian} the CPT invariance of noncommutative  
field theories was studied, and it was
concluded that the CPT theorem holds, both in the case of space-space and
time-space noncommutativity. However, this analysis deals with the  
tree-level action
and it is not sensitive to possible problems arising from quantum  
corrections. In the case of noncommutative
field theories  problems might appear in the form of unitarity  
violations or UV/IR mixing. In particular,
the mixing of scales may jeopardize the tempered nature of the  
Wightman functions as distributions\footnote{A rigorous definition of the class
of functions on which a noncommutative field theory should be built has been given 
in \cite{schwarz}.}. These
issues might demand a revision of the proof of the CPT theorem presented in  
\cite{Chaichian}.

In the present paper we will study in detail some general properties  
of noncommutative quantum field theories,
such as the CPT theorem and the spin-statistics connection, as well as  
the possibility of constructing theories
that are well-defined in the infrared. We will propose an axiomatic  
formulation leading to a proof of the CPT theorem along the lines of the one given  
by Jost for ordinary  theories
\cite{Jost:CPT,Streater:vi}.
In this axiomatic formulation  the vacuum expectation value
of the Heisenberg fields should define tempered distributions. This  
mathematical condition
imposes non-trivial restrictions on the correlation functions 
both in the ultraviolet and
the infrared. In order to give meaning to noncommutative gauge  
theories in the infrared we
present a detailed analysis of noncommutative U(1)$_{\star}$ gauge  
theory with $\mathcal{N}=4$ supersymmetry softly
broken to $\mathcal{N}=0$ in a way that preserves finiteness in the  
ultraviolet and that leads to
a well-defined theory at low energies. The resulting theory provides  
correlation functions that behave
like tempered distributions, while in the infrared we recover the free  
Maxwell theory.
We show that the theory defined in this way satisfies the proposed  
axioms and assess its phenomenological viability.

This paper is organized as follows: in Section \ref{review} we briefly  
review the Seiberg-Witten limit,
and analyze heuristically when CPT-invariance and unitarity  
of the $S$-matrix are expected to hold.
In Section \ref{axiomatics} we adapt the Wightman axioms  
to the noncommutative context. In particular we analyze the issue  
of microscopic causality
\cite{Jost:CPT}. In
Section \ref{cpt} we show that with the modified axioms it is still  
possible to prove the CPT theorem
but, in the general case, the connection between spin and statistics  
is not guaranteed.
In Section \ref{phenomenology} we analyze the softly broken  
$\mathcal{N}=4$ noncommutative U(1)$_\star$ gauge
theory and show that it is possible to formulate it in such a way that  
the adapted axioms are satisfied
at least in perturbation theory. We also make
some general remarks on the possible phenomenological perspectives of  
noncommutative field theories.
Finally, our conclusions are presented in Section \ref{conclusions}.

A final remark is in order before closing this  introduction. We are  
going to study the class of noncommutative
field theories obtained from the Seiberg-Witten limit of string  
theories in the presence of a constant
$B$-field. These theories are defined quantum-mechanically in perturbation theory
in terms  
of their Feynman rules which
follow from the parent string theory. For reasons to be explained in  
the next section, we consider here only
the case of space-space noncommutativity. However,
from a purely field-theoretical point of view,  other  
procedures  can be envisaged to quantize
them. In particular, the authors of Ref. \cite{hamburg},  
motivated by the work of \cite{hamburg:CMP}, proposed
a different way to look at noncommutative theories that leads to a unitary  
$S$-matrix and, if  uncertainty
relations for the space-time coordinates are implemented, they claim  
to obtain not only a unitary theory but also
one that is ultraviolet-finite. In \cite{Sibold} the authors start  
with Dyson's formula to define the
Green functions, and by a careful analysis of the  
passage from time-ordered products to
Wick products, they conclude that the Feynman rules change when there  
is time-space noncommutativity in a way
that preserves perturbative unitarity. For space-space  
noncommutativity the Feynman rules are the
same as those obtained from string theory via the Seiberg-Witten limit.

\section{Heuristic considerations}
\label{review}
\setcounter{equation}{0}

Noncommutative field theories, i.e. theories with ordinary  
products replaced by Moyal products, are
from a quantum point of view theories of dipoles  
\cite{Bigatti:1999iz}. Since the elementary excitations
are extended objects, the resulting theory is nonlocal and the scale  
of non-locality is set by the
length of these dipoles.

This fact is easy to visualize in the cases where noncommutative  
theories arise as effective description of
the dynamics in a certain limit.
In Ref. \cite{SW} it was shown how noncommutative field theories are  
obtained as a particular low-energy limit
of open string theory on D-brane backgrounds in the presence of  
constant NS-NS $B$-field. In this case,
the endpoints of the open strings behave as electric charges in the  
presence of an external magnetic field $B_{\mu\nu}$
resulting in a polarization of the open strings. Labelling by  
$i=1,\ldots,p$  the D-brane
directions, the difference between the zero modes of the string
endpoints 
is given by ($B_{0i}=0$)
\cite{b-field}
\begin{eqnarray}
\Delta X^{i}=X^{i}(\tau,0)-X^{i}(\tau,\pi)=(2\pi\alpha')^2 g^{ij}B_{jk}p^{k},
\label{sw}
\end{eqnarray}
where $g_{\mu\nu}$ is the closed string ($\sigma$-model) metric and  
$p^{\mu}$ is the momentum of the string.
In the ordinary low-energy limit ($\alpha'\rightarrow 0$,  
$g_{\mu\nu}$, $B_{\mu\nu}$  fixed) the distance $|\Delta X|$
goes to zero and the effective dynamics is described by a theory of  
particles, i.e. by a commutative quantum field
theory.

There are, however, other possibilities of decoupling the massive  
modes without collapsing at the same time the length of the
open strings. Seiberg and Witten proposed to consider a low energy  
limit $\alpha'\rightarrow 0$ where both $B_{ij}$ and
the open string metric
\begin{eqnarray}
G_{ij}=-(2\pi\alpha')^2 (Bg^{-1}B)_{ij}
\end{eqnarray}
are kept fixed.
Introducing the notation $\theta^{ij}=(B^{-1})^{ij}$, the separation  
between the string endpoints  can be expressed
as
\begin{eqnarray}
\Delta X^{i}=\theta^{ij}G_{jk}p^{k},
\end{eqnarray}
fixed in the low energy limit. The resulting low-energy  
effective theory is a noncommutative
field theory with noncommutative parameter\footnote{Starting with type  
IIA/IIB string theory in the
presence of $N$ coincident
D$p$-branes, the resulting theory after the Seiberg-Witten limit 
is a $(p+1)$-dimensional
U($N$)$_{\star}$ noncommutative super-Yang-Mills theory with 16  
supercharges.}  $\theta^{ij}$.

The Seiberg-Witten limit can be viewed as a stringy analog of the  
projection onto the lowest Landau level in
a system of electrons in a magnetic field. Indeed, the condition   
fixing the open string metric $G_{\mu\nu}$ 
in the low-energy limit is equivalent to considering a large $B$-field  
limit in which the interaction of the
external field with the end-points of the string dominates the  
worldsheet dynamics. In more physical term,
Eq. (\ref{sw}) can be interpreted as a competition between the Lorentz  
force exerted on the endpoints of the
string and the tension that tends to collapse the string to zero size.  
The Seiberg-Witten limit corresponds
to making the string rigid by taking the tension to infinity (thus  
decoupling the excited states) while at the same time
keeping the $B$-field large in string units, in such a way that the  
length of the string is kept constant.

The previous analysis was confined to  
situations in which the $B_{0i}$ components are set to zero. This results in  
a noncommutative
field theory with only space-space noncommutativity. From a purely  
field-theoretical point of view it is possible
to consider also noncommutative theories where the time coordinate  
does not commute with the spatial ones, i.e.
$\theta^{i0}\neq 0$. In this case, however, non-locality  
is accompanied by a breakdown of unitarity reflected
in the fact that the optical theorem is not satisfied  
\cite{unitarity,AGBZ}. In addition there is no well-defined
Hamiltonian formalism (see, however, \cite{hamburg}).

Noncommutative field theories with time-space noncommutativity cannot be obtained from string theory
in a limit in which strings are completely decoupled. One can try to obtain these kind of theories
by looking for a decoupling limit of string theory in D-brane backgrounds in the presence of $B$-fields with 
$B_{0i}\neq 0$. That the existence of such a decoupling limit is not inmediate can be envisaged 
by noticing that having $B_{0i}\neq 0$ 
is equivalent to a nonvanishing electric field on the brane, and that in this case 
the electric field has to be smaller that the critical value \cite{BP}
\begin{eqnarray}
E<E_{\rm crit}={1\over 2\pi\alpha'}.
\label{bound}
\end{eqnarray}
For fields larger than $E_{\rm crit}$ the background decays via the formation of pairs of open strings in a 
stringy version of the Schwinger mechanism. A partial decoupling limit can be achieved in the limit 
$E\rightarrow E_{\rm crit}$ while keeping $\theta^{\mu\nu}$ and the string tension fixed. 
In this limit closed strings decople from open strings while the latter ones remain in the spectrum \cite{NCOS}. 
Actually, 
the optical theorem can be formally restored by taking into  
account the interchange of undecoupled string states in intermediate channels, 
although they are both tachyonic and have negative norm 
\cite{AGBZ}.

In order to avoid these problems, in the following we  
restrict ourselves to theories with
space-space noncommutativity. Then the existence  
of a parent string theory  provides a good  
guiding principle to study their properties.
The parent string theory also provides a ultraviolet completion  where
non-locality is realized by string fuzziness.

The CPT theorem in string theory has been  
investigated by several groups \cite{CPTstring}.
If the parent string theory satisfies the CPT theorem in
perturbation theory, since
the constant background $B$-field is CPT-even, it is reasonable to  
expect that the noncommutative quantum field
theory obtained in the Seiberg-Witten limit should also preserve CPT.  
At the level of the noncommutative field
theory it is also expected to have CPT-invariance for theories with  
$\theta^{0i}=0$. As discussed above,
these kind of noncommutative theories preserve perturbative unitarity.  
In ordinary quantum field theory there is
an intimate connection between unitarity and CPT-invariance. Indeed,  
if the condition of asymptotic completeness
holds, it can be seen using either the Yang-Feldman relation   
\cite{Yang-Feldman} or the Haag-Ruelle
scattering theory \cite{Haag-Ruelle} that the $S$-matrix can be written in  
terms of the CPT operator of the complete
theory $\Theta$ and the corresponding one for the asymptotic theory  
$\Theta_{0}$ (see also \cite{Bogolubov})
\begin{eqnarray}
S=\Theta^{-1}\Theta_{0}.
\end{eqnarray}
The unitarity of the $S$-matrix follows then from the antiunitarity of  
$\Theta$ and $\Theta_{0}$. Therefore a theory
with CPT invariance is likely to be unitary.

Motivated by these heuristic arguments we proceed to construct the CPT theorem 
along the lines of the
proof given in \cite{Streater:vi}. As a first  
step we attempt an axiomatic formulation
of noncommutative field theories as a modification of the  
Wightman axioms \cite{Wightman,books,Haag:book}.

\section{Modified constructions}
\label{axiomatics}
\setcounter{equation}{0}

Compared to ordinary theories, two basic features of  
noncommutative  theories are their non-locality and
the breaking of Lorentz invariance due to the presence of the  
antisymmetric tensor $\theta^{\mu\nu}$. In the following we
will take it of the form
\begin{eqnarray}
\theta^{\mu\nu}=\left(
\begin{array}{cccc}
0 & \theta_{e} & 0 & 0 \\
-\theta_{e} & 0 & 0 & 0 \\
0 & 0 & 0 & \theta_{m} \\
0 & 0 & -\theta_{m} & 0
\end{array}
\right),
\label{theta}
\end{eqnarray}
with $\theta_{e},\theta_{m}\in\mathbb{R}$.
The Wightman axioms have to be modified in order to  
accommodate noncommutativity and reduced Poincar\'e invariance. Eventually we will
be interested in theories with space-space noncommutativity  
$\theta_{e}=0$. Nevertheless, for most of this section we will keep
$\theta_{e}$ explicitly.

\subsection{Symmetries of noncommutative field theories}

The commutation relations for the coordinates  
$[x^{\mu},x^{\nu}]=i\theta^{\mu\nu}$ are not preserved
by the  Lorentz group O(1,3) although they are indeed  
left invariant by the group of rigid translations
$x^{\mu}\rightarrow x^{\mu}+a^{\mu}$ with  
$a^{\mu}\in\mathbb{R}$. Using\footnote{For a general $\theta^{\mu\nu}$
the unbroken subgroup of O(1,3) is the one preserving 
both $\theta^{0i}$ and $\epsilon_{ijk}\theta^{jk}$. Hence, in general, 
the Lorentz group is completely broken unless it has the form of Eq. (\ref{theta}).}
(\ref{theta}) it is
straightforward to find that the largest subgroup of O(1,3) leaving  
invariant the commutation relations is SO(1,1)$\times$SO(2),
where the SO(1,1) factor acts on the ``electrical'' coordinates  
$x_{e}=(x^{0},x^{1})$ whereas SO(2) rotates the ``magnetic'' ones
$\vec{x}_{m}=(x^{2},x^{3})$. In the case of space-space noncommutativity 
($\theta_{e}=0$) this group is enhanced to O(1,1)$\times$ SO(2). Since 
we will be mostly concerned about this latter case, we will take the symmetry group of the
noncommutative theory to be  
$\mathfrak{P}=[\mbox{O(1,1)}\times\mbox{SO(2)}]\rtimes  
\mathcal{T}_{4}$,where $\mathcal{T}_{4}$ is the group of translations.

The representation theory of O(1,1)$\times$SO(2) shares many common  
features with that of the Lorentz group
O(1,3). Representations of SO(2) are parametrized by the angle  
$\varphi\in[0,2\pi]$.
The O(1,1) factor has a richer structure. As  
in the standard case, it has four
sheets
\begin{eqnarray}
&\mathfrak{L}_{\pm}^{\uparrow}:&\hspace*{0.5cm} \det{\Lambda}=\pm 1;  \hspace*{0.25cm}
\Lambda^{0}_{\,\,0}\geq 1, \\
&\mathfrak{L}_{\pm}^{\downarrow}:&\hspace*{0.5cm} \det{\Lambda}=\pm 1;  \hspace*{0.25cm}
\Lambda^{0}_{\,\,0}\leq -1.
\end{eqnarray}
The elements of the four sheets can be expressed as a  
product of an element of $\mathfrak{L}_{+}^{\uparrow}$
and one of the three discrete transformations inverting one or the two  
electric coordinates
\begin{eqnarray}
\mathfrak{L}_{+}^{\downarrow}=\mathfrak{L}_{+}^{\uparrow}I_{\rm st},  
\hspace*{1cm}
\mathfrak{L}_{-}^{\uparrow}=\mathfrak{L}_{+}^{\uparrow}I_{\rm s},  
\hspace*{1cm}
\mathfrak{L}_{-}^{\downarrow}=\mathfrak{L}_{+}^{\uparrow}I_{\rm t}
\end{eqnarray}
where
\begin{eqnarray}
I_{\rm st}=\left(
\begin{array}{cc}
-1 & 0 \\
0 & -1
\end{array}
\right), \hspace*{1cm}
I_{\rm s}=\left(
\begin{array}{cc}
1 & 0 \\
0 & -1
\end{array}
\right), \hspace*{1cm}
I_{\rm t}=\left(
\begin{array}{cc}
-1 & 0 \\
0 & 1
\end{array}
\right).
\label{st}
\end{eqnarray}

Each element of $\mathfrak{L}_{+}^{\uparrow}$ is parametrized by  
a single number $\gamma\in\mathbb{R}$,
the rapidity of the boosts along $x^{1}$
\begin{eqnarray}
\Lambda=\left(
\begin{array}{cc}
\cosh\gamma & \sinh\gamma \\
\sinh\gamma & \cosh\gamma
\end{array}
\right).
\end{eqnarray}
The transformations of $x_{e}$ are specially simple
introducing null-coordinates $x^{\pm}={1\over\sqrt{2}}(x^{0}\pm  
x^{1})$. These transform according to
\begin{eqnarray}
\mathfrak{L}_{+}^{\uparrow}:x^{\pm}\longrightarrow e^{\pm\gamma}x^{\pm}
\end{eqnarray}
together with
\begin{eqnarray}
I_{\rm st}x^{\pm}=-x^{\pm}, \hspace*{1cm}
I_{\rm s}x^{\pm}=x^{\mp}, \hspace*{1cm}
I_{\rm t}x^{\pm}=-x^{\mp}.
\end{eqnarray}

Similar to four dimensions,
the complexification of O(1,1)  connects continuously different  
sheets. In particular
$\mathfrak{L}_{+}^{\uparrow}$ and $\mathfrak{L}_{+}^{\downarrow}$ can  
be connected
by a family of transformations with complex rapidity $\gamma+i\vartheta$ where
$0\leq \vartheta\leq \pi$. Denoting  
$\mathfrak{L}_{+}=\mathfrak{L}_{+}^{\uparrow}\cup\mathfrak{L}_{+}^{\downarrow}\equiv \mbox{SO(1,3)}$
we find
\begin{eqnarray}
\mathfrak{L}_{+}(\mathbb{C})=\mathbb{C}^{*}\equiv \mathbb{C}-\{0\}
\end{eqnarray}
where the action of $\mathfrak{L}_{+}(\mathbb{C})$ on the coordinates  
is given by $x^{\pm}\rightarrow z x^{\pm}$,
$z\in\mathbb{C}^{*}$. In particular, this result implies that the  
transformation inverting $x_{e}$ can be continuously
connected with the identity using transformations of  
$\mathfrak{L}_{+}(\mathbb{C})$. At the same time
$\vec{x}_{m}$ can also be inverted by a 180 degrees SO(2)-rotation. This
result is central in the proof of the CPT theorem (see Section \ref{cpt}).

Representations of 
SO(1,1) are labelled by a helicity $\lambda\in\mathbb{R}$ and those of
SO(2) by the ``angular momentum'' $j$ ($2j\in\mathbb{Z}$). Any quantity $\Psi^{(j)}_{\lambda}$  
in the $(\lambda,j)$ representation of O(1,1)$\times $ SO(2) transforms 
as
\begin{eqnarray}
\Psi^{(j)}_{\lambda}\longrightarrow e^{\lambda\gamma}\,e^{j\varphi}\,\Psi^{(j)}_{\lambda}.
\end{eqnarray}
In $\mathfrak{P}$ we can build two Casimir operators with $P_{\mu}$  
\begin{eqnarray}
s \equiv (P^{0})^2-(P^{1})^2, \hspace*{1cm}
t \equiv (P^{2})^{2}+(P^{3})^2,
\end{eqnarray}
In perturbative  
calculations in noncommutative field theories
these two invariants appear usually in the combinations
\begin{eqnarray}
p^2=s-t, \hspace*{1cm} p\circ p=\theta_{e}^2 s+\theta_{m}^2 t.
\label{p2}
\end{eqnarray}

Since by definition $t\geq 0$, the representations of $\mathfrak{P}$  
can be classified according to the sign of
$s$ into ``massless'' ($s=0$), ``massive'' ($s>0$) and ``tachyonic''  
($s<0$). Since $\mathfrak{P}\subset
\mathscr{P}={\rm O}(1,3)\rtimes \mathcal{T}_{4}$,   
 it is important to
notice that the type of the representations of $\mathfrak{P}$ do not  
necessarily coincide with those of $\mathscr{P}$. In particular, from (\ref{p2}),
we see that ``massive'' and ``massless'' representations of  
$\mathfrak{P}$ can
be tachyonic when interpreted as representations of $\mathscr{P}$. On  
the other hand, any ``tachyonic'' representation
of $\mathfrak{P}$ will be tachyonic with respect to $\mathscr{P}$, but  
not the other way around.

\subsection{Microcausality}
\label{microcausality}

The second issue that has to be studied in our attempt to find an  
axiomatic formulation of noncommutative field theories
is the notion of microcausality. In ordinary  theories  
causality is implemented  by
demanding that fields should commute or anticommute outside their  
relative light cone. For noncommutative 
theories there is no structure preserving the  
four-dimensional Lorentz group O(1,3) and the concept
of a light cone is lost.

From our discussion of the symmetries of noncommutative field theories  
it follows that causality, if preserved at all,
must be defined with respect to the O(1,1) factor of the symmetry  
group. This means that the light cone $x^2=0$ is actually replaced by
the ``light wedge'' $V_{+}=\{x\in \mathbb{R}^{1,3}|x_{e}^2=0\}$ (see Fig. 1).
\begin{figure}[ht]
\centerline{ \epsfxsize=4.0truein \epsfbox{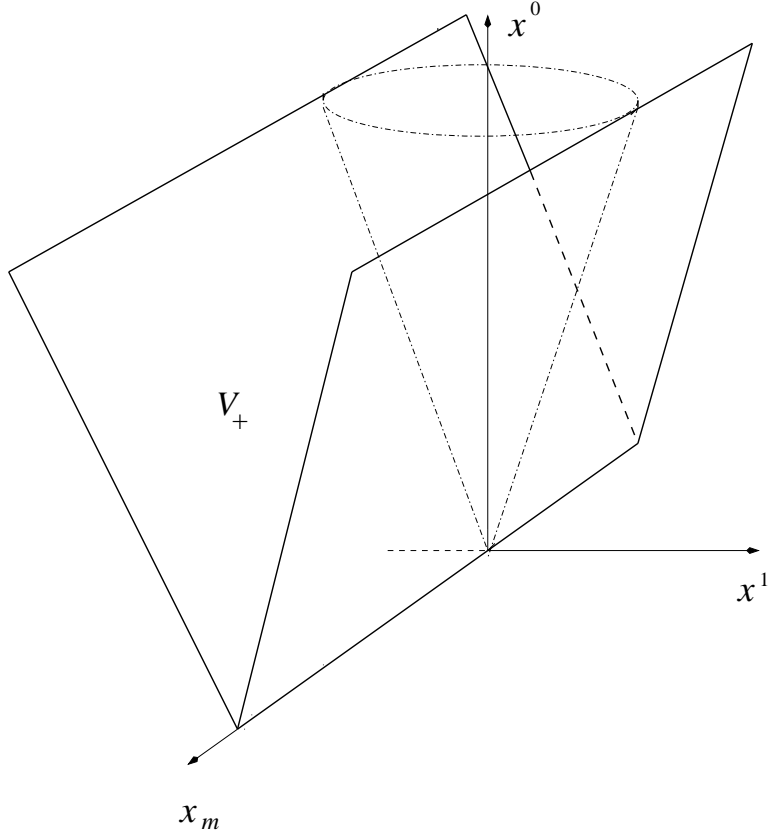}}
\caption{The causal wedge $V_{+}$.}
\end{figure}
This suggests a change in the concept of microcausality by replacing  
the light cone by the light wedge, so fields will either commute
or anticommute whenever their relative coordinate $x-y$ satisfies  
$(x_e-y_e)^2<0$.

In the following, we study a weaker version of this condition,  
namely under which conditions the vacuum expectation value
of the commutator of two noncommutative scalar fields vanishes outside  
the light wedge. To achieve this we construct a
K\"all\'en-Lehmann representation for $\langle  
\Omega|[\phi(x),\phi(y)]|\Omega\rangle$ (with $|\Omega\rangle$ the  
unique vacuum
state of the theory) using  invariance  under the group
$\mathfrak{P}$. Therefore we introduce the  
[O(1,1)$\times$SO(2)]-invariant measure:
\begin{eqnarray}
d\mu(p)={d^{4}p\over (2\pi)^4}  
\vartheta(p^0)\,(2\pi)\,\delta(p_{e}^2-\alpha^2)\,(2\pi)\,\delta(\vec{p}_{m}^{\,2}-\beta^{2}).
\end{eqnarray}
Using a basis of the Hilbert space $\{|j,p_e,\vec{p}_{m}\rangle\}$,  
where $j$ is a collective index denoting
any other set of quantum numbers
needed to specify the state, we write the closure relation
\begin{eqnarray}
1=|\Omega\rangle\langle\Omega|+\sum_{j}\int  
d\mu(p_j)|j,p_e,\vec{p}_{m}\rangle \langle j,p_e,\vec{p}_{m}|.
\label{identity}
\end{eqnarray}
Starting with $\langle\Omega|\phi(x)\phi(y)|\Omega\rangle$ and  
inserting the identity as in  (\ref{identity}) one arrives,
after some algebra, at
\begin{eqnarray}
\langle\Omega|[\phi(x),\phi(y)]|\Omega\rangle &=&i\, {\rm  
sign}(x^{0}-y^{0})
\label{kl}\\
&\times& \int_{-\infty}^{\infty}{d\alpha^2\over 2\pi}
\int_{0}^{\infty}{d\beta^2\over 2\pi} \varrho(\alpha^2,\beta^2) {\rm Im\,}
G_{F}(x_{e}-y_{e},\alpha^2)\,J_{0}\left(\sqrt{\beta^2(\vec{x}_{m}-\vec{y}_{m})^{2}}\,\right).
\nonumber
\end{eqnarray}
Here $J_{0}(x)$ is a Bessel function of the first kind, and $G_{F}(x,m^2)$ is the
two-dimensional Feynman propagator for a free scalar field of mass $m$,
\begin{eqnarray}
G_{F}(x_{e}-y_{e},m^2)=\int {d^2p_{e}\over (2\pi)^2}{i\over  
p_{e}^2-m^2+i\epsilon}e^{-ip_{e}\cdot (x_{e}-y_{e})}.
\end{eqnarray}
Finally, the spectral function $\varrho(\alpha^2,\beta^2)$ is given by
\begin{eqnarray}
\varrho(\alpha^2,\beta^2)=\sum_{j}(2\pi)\,\delta(\alpha^2-\alpha_{j}^2)\,
(2\pi)\,\delta(\beta^2-\beta_{j}^2)\,\left|\langle\Omega|\phi(0)|j,p_e,\vec{p}_{m}\rangle\right|^2.
\label{specdens}
\end{eqnarray}
Using the invariance of the theory under $\mathfrak{P}$ it is  
straightforward to show that the overlaps
$\left|\langle\Omega|\phi(0)|j,p_e,\vec{p}_{m}\rangle\right|^2$ only  
depend on $\alpha_{j}^2$ and $\beta_{j}^2$,
the eigenvalues of the Casimir  operators $P_{e}^2$ and  
$\vec{P}_{m}^{\,2}$ corresponding to the state $|j,p_e,\vec{p}_{m}\rangle$.

Up to this point we have only used the symmetry properties of the  
theory, without any reference to the fact that we might
be dealing with a noncommutative theory. Actually, from Eqs.  
(\ref{kl}) and (\ref{specdens}) it is possible to extract some
general consequences. Taking into account the property of the  
two-dimensional Feynman propagator
\begin{eqnarray}
{\rm Im\,}G_{F}(x_{e}-y_{e},m^2)=0, \hspace*{1cm} \mbox{for} \,\,\,\,  
(x_{e}-y_{e})^2<0 \,\,\,\, \mbox{and} \,\,\,\, m^{2}\geq 0
\end{eqnarray}
we conclude that the vacuum expectation value of the commutator  
vanishes outside the light wedge whenever
the theory does not contain in its Hilbert space tachyonic  
representations of O(1,1).

If there are  O(1,1)-tachyonic states  in the spectrum, i.e. if the spectral  
function $\varrho(\alpha^2,\beta^2)$ has support
for $\alpha^2<0$,  the imaginary part of the Feynman propagator does  
not vanish and
in general $\langle\Omega|[\phi(x),\phi(y)]|\Omega\rangle$ will  
not be zero outside the light wedge.
In the latter case, the modification of microcausality proposed here  
will not be satisfied (cf.  \cite{AGBZ}).

Particularizing this result to the case of noncommutative field  
theories we conclude that, in general, the
adapted notion of microcausality is not preserved by theories with  
time-space noncommutativity.
For them poles in the two-point function at $p\circ p<0$ 
induce a support of $\varrho(\alpha^2,
\beta^2)$ for $\alpha^2<0$. This also occurs for theories with  
space-space noncommutativity but containing
O(1,1)-tachyons, as it is the case for noncommutative QED  
\cite{QEDtachyon}. In Section \ref{phenomenology}
we will study how to construct theories in which this tachyonic  
instabilities are absent and look like QED at low energies.

\subsection{The adapted axioms}
\label{adapted}

After discussing the consequences of  Lorentz symmetry breaking 
in noncommutative theories,
we come to the modification of Wightman axioms needed
to accommodate noncommutative theories.  Here we will not discuss in  
details the whole set of axioms \cite{Streater:vi}
but only comment on the modifications required. 

Concerning the space of states of the theory, we take it to  
be given by a (separable) Hilbert space carrying a
unitary representation of the group\footnote{When gauge fields are present, 
we have to generalize this condition to admit an indefinite Hilbert space. This permits
a rigorous implementation of the Gupta-Bleuler procedure \cite{strocchi-wightman}. For 
simplicity, however, we do not consider this more general situation.}
$\mathfrak{P}=[\mbox{O(1,1)}\times\mbox{SO(2)}]\rtimes  
\mathcal{T}_{4}$. In addition
we will assume the existence of a unique vacuum state $|\Omega\rangle$  
invariant under $\mathfrak{P}$.
The spectrum of the momentum operator $P$ is in the forward  
light wedge
\begin{eqnarray}
\mbox{Spec}(P)=\left\{(p^0)^2-(p^1)^2\geq 0, p^0\geq 0\right\}.
\end{eqnarray}

Fields $\Phi(x)$ are operator-valued distributions in the  
noncommutative space $\mathbb{R}^{1,3}$
transforming under $\mathfrak{P}$ as
\begin{eqnarray}
\mathcal{U}(\Lambda,a)\,\Phi(x)\,\mathcal{U}(\Lambda,a)^{-1}={U}_\Lambda\Phi(\Lambda  
x+a)
\end{eqnarray}
where $U_{\Lambda}$ is a matrix acting on the indices of the field $\Phi(x)$.
Wightman functions, i.e. vacuum expectation values of fields,
\begin{eqnarray}
W(x_{1},\ldots,x_{n})\equiv  
\langle\Omega|\Phi(x_{1})\ldots\Phi(x_{n})|\Omega\rangle
\end{eqnarray}
define tempered distributions on the Schwartz space  
$\mathscr{S}[(\mathbb{R}^{1,3})^{n}]$ of smooth test functions which  
decrease, together with
all their derivatives, faster than any power at infinity  
\cite{schwartz}. As we will
see in Section \ref{phenomenology},
in noncommutative theories, the tempered character of the  
distributions can be spoiled by the appearance of hard infrared
singularities induced by quadratic divergences due to UV/IR mixing.

Finally,  as argued above, in order to adapt the postulate of  
microcausality one has to relax the condition that
field (anti)commutators vanish outside the light cone. This is done by  
replacing the light cone by the light wedge, namely
\begin{eqnarray}
[\Phi(x),\Phi(y)]_{\pm}=0 \hspace*{0.5cm} \mbox{if} \hspace*{0.5cm}  
(x^0-y^0)^2-(x^1-y^1)^2<0.
\label{LC}
\end{eqnarray}
The rest of the axioms, such as hermiticity and completeness are taken over   
without modification.

\section{The CPT theorem in noncommutative quantum field theory}
\label{cpt}
\setcounter{equation}{0}

In what follows we proceed to prove the CPT theorem for  
noncommutative theories satisfying the adapted axioms
of Section \ref{adapted}. In order to keep things simple,  
we restrict our analysis in the case of a
real scalar field. The generalization to fields transforming in other  
representations of O(1,1)$\times$SO(2) is not difficult.

In ordinary theories satisfying the  
Wightman axioms, the CPT theorem can be proved directly on the
Wightman functions by taking advantage of their analyticity properties  
\cite{Jost:CPT}. Because of translational invariance,
the $n$-point Wightman function $W_{n}(x_{1},\ldots,x_{n})$ is  
actually a function of the $(n-1)$ coordinate differences
$\xi_{1}=x_1-x_2,\ldots,\xi_{n-1}=x_{n-1}-x_n$
\begin{eqnarray}
W_{n}(x_1,\ldots,x_n)=\mathscr{W}_{n}(x_1-x_2,\ldots,x_{n-1}-x_{n})\equiv  
\mathscr{W}_{n}(\xi_{1},\ldots,\xi_{n-1}).
\end{eqnarray}
The functions $\mathscr{W}_{n}(\xi_{1},\ldots,\xi_{n-1})$ can be  
analytically continued into the tube $\mathscr{T}_{n-1}=
\{\xi_{j}-i\eta_{j}, (\eta_{j})^2>0,\eta_j^0>0\}$, and further into  
the extended tube $\mathscr{T}_{n-1}'$
containing all the points in $\mathbb{C}^{4(n-1)}$
that can be reached from $\mathscr{T}_{n-1}$ by a complex Lorentz  
transformation.
The CPT theorem is proved by noticing that the analytic continuation  
of the Wightman function $\mathscr{W}_{n}(\xi_{1},\ldots,\xi_{n-1})$
and its CPT-transformed $\mathscr{W}_{n}(\xi_{n-1},\ldots,\xi_{1})$  
coincide on a real neighborhood of $\mathscr{T}_{n-1}'$, so
by the ``edge of the wedge'' theorem they have to coincide on their  
whole domain of analyticity. A key ingredient in the proof is that
at spatially separated real points the following identity holds
\begin{eqnarray}
\mathscr{W}_{n}(-\xi_{n-1},\ldots,-\xi_1)=\mathscr{W}_{n}(\xi_{1},\ldots,\xi_{n-1}),  
\,\,
\mbox{for real $(\xi_{1},\ldots,\xi_{n-1})\in \mathscr{T}_{n-1}'$
and $\xi_j^2<0$}.
\end{eqnarray}
This property, weak local commutativity, follows from the postulate of  
microcausality.

\subsection{Preliminaries}

For noncommutative theories, we follow the strategy 
outlined above. 
Two important ingredients  
in the proof are: the possibility of
implementing the
PT-transformation on the coordinates by a transformation of the  
complexified symmetry group $\mathfrak{L}_{+}(\mathbb{C})$,
and a weaker version of local commutativity (\ref{LC}). Since the  
causal structure of the theory is fully determined by the
O(1,1)  symmetry, we  
perform analytic continuation of the Wightman function
{\sl only} with respect to the ``electrical'' coordinates  
$x_e=(x^0,x^1)$, while the ``magnetic'' ones $\vec{x}_{m}=(x^2,x^3)$ are  
left as
spectators, keeping in mind that the inversion $\vec{x}_m\rightarrow  
-\vec{x}_{m}$ is implemented by a 180 degrees SO(2) rotation.

We consider the $n$-point Wightman function for a noncommutative  
scalar field theory\footnote{Notice that, since the
noncommutative theory is invariant under
the group of translations, the Wightman function only depends on the  
$(n-1)$ coordinate differences as in the
commutative case.} $\mathscr{W}_{n}(\xi_{1},\ldots,\xi_{n})$. Using  
arguments parallelling the standard case
(see, for example, \cite{Haag:book}) the  
Wightman functions can be analytically continued in the
electric coordinates to the tube
\begin{eqnarray}
\mathfrak{T}_{n-1}=\Big\{\xi_{j}-i\eta_{j}\Big| (\eta_{j})_{e}^2>0,  
\eta_j^0>0, (\vec{\eta}_{j})_{m}=0\Big\}.
\end{eqnarray}
By definition, the set $\mathfrak{T}_{n-1}$ does not contain any real  
point. The original Wightman function
$\mathscr{W}_{n}(\xi_{1},\ldots,\xi_{n})$
is the boundary value of the analytic function defined on this tube.  
However, similarly to the commutative case,
the Wightman function can be further analytically continued to the  
so-called extended tube $\mathfrak{T}_{n-1}'$ formed by all the
points reachable from $\mathfrak{T}_{n-1}$ by a  
transformation in $\mathfrak{L}_{+}(\mathbb{C})$. It is important to
notice that the transformations of $\mathfrak{L}_{+}(\mathbb{C})$  
leave the magnetic coordinates invariant, so we never
perform any analytical continuation on them. This is one of the main  
difference with respect to the commutative case where the  
continuation is made in all four coordinates.

We can use the very definition of the extended tube  
$\mathfrak{T}_{n-1}'$ to analytically continue the Wightman  
functions
on it. If $(\zeta_1,\ldots,\zeta_{n-1})\in\mathfrak{T}_{n-1}'$,  
by definition, there exists a transformation
$\Lambda\in\mathfrak{L}_{+}
(\mathbb{C})$ such that  
$(\Lambda\zeta_1,\ldots,\Lambda\zeta_{n-1})\in\mathfrak{T}_{n-1}$. The  
value of the Wightman
function in $(\zeta_1,\ldots,\zeta_{n-1})$ is given by\footnote{In the  
general case where fields in arbitrary representations
of $\mbox{O(1,1)}\times\mbox{SO(2)}$ are considered,  the Wightman  
functions transform covariantly with respect to the
transformations in $\mathfrak{L}_{+}(\mathbb{C})$. In this case, there  
would be nontrivial factors on the right-hand side of Eq. (\ref{def})
depending on the helicities of the fields involved.}
\begin{eqnarray}
\mathscr{W}_{n}(\zeta_1,\ldots,\zeta_{n-1})\equiv  
\mathscr{W}_{n}(\Lambda\zeta_1,\ldots,\Lambda\zeta_{n-1}).
\label{def}
\end{eqnarray}

The condition to have a unique analytical continuation to  
$\mathfrak{T}_{n-1}'$ requires to verify that the definition (\ref{def}) is  
actually
unique: if the same point  
$(\zeta_1,\ldots,\zeta_{n-1})\in\mathfrak{T}_{n-1}'$ can be reached  
from two different
points  
$(\omega_1,\ldots,\omega_{n-1}),(\rho_{1},\ldots,\rho_{n-1})\in\mathfrak{T}_{n-1}$  
by means of the transformations
$\Lambda,\Lambda'\in\mathfrak{L}_{+}(\mathbb{C})$, Eq. (\ref{def})  
should yield the same value for the Wigthman function
$\mathscr{W}_{n}(\zeta_{1},\ldots,\zeta_{n-1})$. In our case this is  
guaranteed by the condition that for all
$(\omega_1,\ldots,\omega_{n-1})\in \mathfrak{T}_{n-1}$ and all  
$\Lambda\in\mathfrak{L}_{+}(\mathbb{C})$ such that
$(\Lambda\omega_1,\ldots,\Lambda\omega_{n-1})$ is also a point of the  
tube $\mathfrak{T}_{n-1}$, it holds that
\begin{eqnarray}
\mathscr{W}_{n}(\omega_1,\ldots,\omega_{n-1})=\mathscr{W}_{n}(\Lambda\omega_1,\ldots,\Lambda\omega_{n-1}).
\label{complex}
\end{eqnarray}
If $\Lambda$ is a real transformation, the condition (\ref{complex})  
obviously holds because of the covariance of the Wightman function.
The way to prove that the condition is satisfied also for arbitrary  
complex transformations is
to perform analytic continuation of  (\ref{complex}) along curves  
in the complex-rapidity plane of O(1,1) boosts.
This works if there exists a one-parameter family of complex  
transformations $\Lambda(t)\in\mathfrak{L}_{+}(\mathbb{C})$,
$0\leq t\leq 1$ interpolating between the identity and the  
transformations $\Lambda$ such that all the points
$(\Lambda(t)\omega_1,\ldots,\Lambda(t)\omega_{n-1})$ lie in the tube  
$\mathfrak{T}_{n-1}$.

Because of the Abelian character of the symmetry group the proof of  
this result is much simpler than in the commutative case. 
We start with $\zeta \in\mathfrak{T}_1$
and choose null  
coordinates $\zeta^{\pm}$. The condition that
this point belongs to the tube implies ${\rm  
Im\,}\zeta^{\pm}<0$.  Under a transformation of
$\mathfrak{L}_{+}(\mathbb{C})$ with complex rapidity $\alpha+i\beta$,  
$\zeta^{\pm}\rightarrow e^{\pm(\alpha+i\beta)}\zeta^{\pm}$. This point  
belong to $\mathfrak{T}_{1}$ if
${\rm Im\,}e^{\pm(\alpha+i\beta)}\zeta^{\pm}<0$. With this setup it is  
easy to prove that there is always a one-parameter
family of complex transformations interpolating between the identity and a 
given complex transformations and such that the whole family lies in the
tube.
\begin{figure}[ht]
\centerline{ \epsfxsize=4.0truein \epsfbox{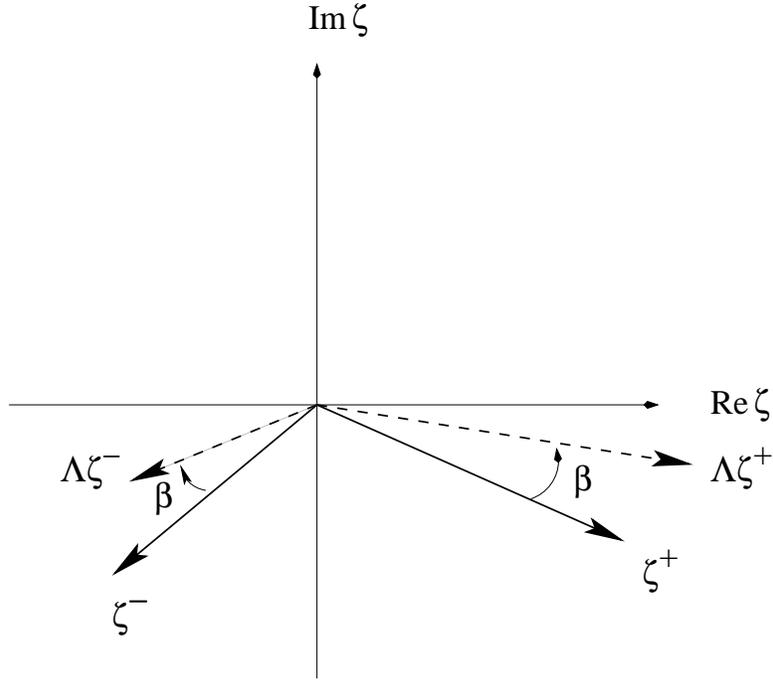}}
\caption{Under a complex O(1,1) boost with rapidity $\alpha+i\beta$  
the null coordinates $\zeta^\pm$ are rotated in
opposite directions. A one-parameter family of transformations can be  
constructed by considering two monotonous functions
$(\alpha(t),\beta(t))$ with $0\leq t\leq 1$
interpolating between (0,0) and $(\alpha,\beta)$ such that all the  
transformed points lie in the lower half-plane
and therefore belong to $\mathfrak{T}_1$.}
\end{figure}
As shown in Fig. 2  the effect of a complex transformation is to  
rotate the null coordinates $\zeta^{\pm}$ by an angle $\beta$ in
opposite directions (at the same time than rescaling them by 
$e^\alpha$). Therefore, if the initial and the
final points lie in $\mathfrak{T}_1$,
i.e. in the lower half plane, there is always a family of  
transformations which interpolate between the two and such that the  
transformed
\begin{figure}[ht]
\centerline{ \epsfxsize=6.5truein \epsfbox{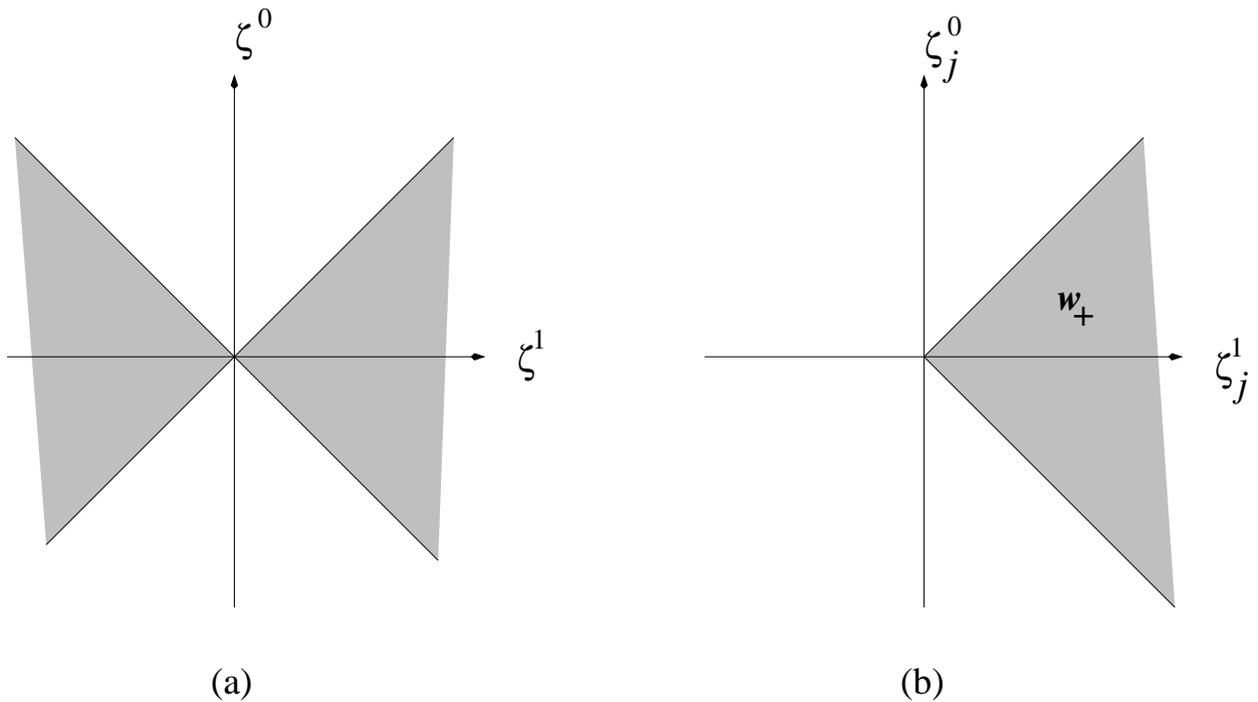}}
\caption{Set of Jost points: (a) for $\mathfrak{T}_1'$; (b) for  
$\mathfrak{T}_{n-1}'$ with $n>2$ the set of Jost points is the
product of $n-1$ copies of the wedge $w_+$.}
\end{figure}
points always have negative imaginary parts. The extension to  
$\mathfrak{T}_{n-1}$ with $n>2$ is straightforward.

This proves that the Wightman functions can be  
analytically continued to the extended tube $\mathfrak{T}_{n-1}'$.
Unlike the tube $\mathfrak{T}_{n-1}$, the extended tube does contain  
real points. By analogy with the standard case, we 
call these real points {\em Jost points}. Actually,  
the analog of a theorem proved by Jost \cite{Jost:CPT}
holds in this case,
namely, that a point $(\zeta_1,\ldots,\zeta_{n-1})\in  
\mathfrak{T}_{n-1}'$ is real if and only if, any combination of the  
form
\begin{eqnarray}
\sum_{j=1}^{n-1}\lambda_j\zeta_j,\,\,\,\, \mbox{with}  
\,\,\,\,\lambda_j\geq 0 \,\,\,\, \mbox{and}\,\,\,\,  
\sum_{j=1}^{n-1}\lambda_j^2>0
\end{eqnarray}
lies outside the light wedge,
\begin{eqnarray}
\left(\sum_{j=1}^{n-1}\lambda_j\zeta_j^0\right)^2-\left(\sum_{j=1}^{n-1}\lambda_j\zeta_j^1\right)^2<0.
\label{jost}
\end{eqnarray}

Although the condition (\ref{jost}) is weaker than the ordinary  
one, the Jost points still form a real
neighborhood
in the extended tube on which the ``edge of the wedge'' theorem can be  
applied. For $\mathfrak{T}_1'$ the set of Jost points is the shadowed region depicted
in Fig. 3a. They are formed by the wedge $(\zeta_{e})^2<0$. On the other  
hand, for $\mathfrak{T}_{n-1}$ with $n>2$ the set of Jost points
is given by $w_+^{n-1}$, where $w_+$ is the wedge defined by  
$(\zeta_e)^2<0$ and $\zeta^1>0$, as shown in Fig. 3b.

\subsection{The proof of the theorem}

We are ready now to proceed to demonstrate the CPT theorem.  
The main difference with the standard proof lies in the fact analytic  
continuation is performed only 
in the electric components of the coordinates
and only invariance under O(1,1)$\times$SO(2) is assumed. However, the  
existence of this unbroken subgroup of the Lorentz group
is enough to ensure that the PT transformation can be connected to the  
identity by a family of complex boosts.

In terms of the Wightman functions, the CPT theorem 
for a neutral scalar field $\phi(x)$ states that
\begin{eqnarray}
\langle\Omega|\phi(x_1)
\ldots\phi(x_{n})|\Omega\rangle =  
\langle\Omega|\phi(-x_{n})\ldots
\phi(-x_{1})|\Omega\rangle.
\label{wcpt}
\end{eqnarray}
This identity is equivalent to the following one for the analytically   
continued  Wightman function into the extended
tube $\mathfrak{T}_{n-1}'$ (cf. \cite{Jost:CPT,Streater:vi})
\begin{eqnarray}
\mathscr{W}_{n}(\zeta_{1},\ldots,\zeta_{n-1})=\mathscr{W}_{n}(\zeta_{n-1},\ldots,\zeta_{1}),  
\hspace*{1cm} (\zeta_{1},\ldots,\zeta_{n-1})\in\mathfrak{T}_{n-1}'.
\label{cpt1}
\end{eqnarray}
We prove now that Eq. (\ref{wcpt}), and therefore the CPT theorem,  
is equivalent to the condition of
adapted weak local commutativity
\begin{eqnarray}
\langle\Omega|\phi(x_1)\ldots\phi(x_{n})|\Omega\rangle =  
\langle\Omega|\phi(x_{n})\ldots\phi(x_{1})|\Omega\rangle  
\hspace*{1cm}
\label{awlc}
\end{eqnarray}
with $(x_{1}-x_{2},\ldots,x_{n-1}-x_{n})$ a Jost point. Since Jost  
points are real and lie outside the light wedge, the condition
(\ref{awlc}) follows from the adapted postulate of microcausality  
introduced in Section \ref{adapted}.

We begin by rewriting the condition (\ref{awlc}) as an
identity between Wightman functions at Jost points
\begin{eqnarray}
\mathscr{W}_{n}(\zeta_1,\ldots,\zeta_{n-1})=\mathscr{W}_{n}(-\zeta_{n-1},\ldots,-\zeta_{1}).
\label{mwlc}
\end{eqnarray}
Since this identity holds at Jost points, and these form a real  
neighborhood of the domain of analyticity of the Wightman functions,  
one
concludes, using the ``edge of the wedge'' theorem \cite{Streater:vi},  
that Eq. (\ref{awlc}) is valid in the whole extended tube
$\mathfrak{T}_{n-1}'$.

As it was discussed in Section \ref{review}, the inversion of the four  
space-time coordinates is the product of the transformation $I_{\rm  
st}\in
\mathfrak{L}_{+}(\mathbb{C})$ [see Eq. (\ref{st})] and a  
SO(2)-rotation of $\pi$. Therefore, the transformation $I_{\rm st}$
is a real transformation belonging to
O(1,1)$\times$SO(2).  
Applying the covariance of the Wightman functions  
under
$\mathfrak{L}_{+}(\mathbb{C})\times \mbox{SO(2)}$ we can  
write\footnote{The invariance of the Wightman function in the
extended tube under $I_{\rm st}$ follows from Eq. (\ref{def}).} 
\begin{eqnarray}
\mathscr{W}_{n}(\zeta_1,\ldots,\zeta_{n-1})=\mathscr{W}_{n}(I_{\rm  
st}\zeta_{n-1},\ldots,I_{\rm st}\zeta_{1})=
\mathscr{W}_{n}(\zeta_{n-1},\ldots,\zeta_{1}).
\end{eqnarray}
This identity is true for all points 
in $\mathfrak{T}_{n-1}'$ and therefore it also holds 
in the tube $\mathfrak{T}_{n-1}$. Since the Wightman function
$\langle\Omega|\phi(x_1)\ldots\phi(x_{n})|\Omega\rangle$
are the boundary values of the analytic function  
$\mathscr{W}_{n}(\zeta_1,\ldots,\zeta_{n-1})$ defined in the tube,
this proves the CPT theorem (\ref{wcpt}) for a neutral scalar field

The reverse is also easily proved. Indeed, if Eq. (\ref{cpt1}) holds  
in the extended tube using again the covariance of the
Wightman functions under  
$\mathfrak{L}_{+}(\mathbb{C})\times\mbox{SO(2)}$ we conclude that Eq.  
(\ref{mwlc}) holds also at all points in
the extended tube. Therefore the relation is satisfied in particular at the  
Jost points and
we recover the adapted weak local commutativity condition (\ref{awlc}).

We have proved that the CPT theorem holds in noncommutative quantum  
field theories satisfying the adapted axioms, in particular
the postulate of weak local commutativity. As discussed in Section  
\ref{microcausality} the vacuum expectation value of
the commutator of two scalar fields does not vanish for those
noncommutative theories containing states in tachyonic representations  
of $\mathfrak{P}$. This implies
that for this type of theories the adapted postulate of microcausality  
does not hold in general.  This is for example the case
of theories with time-space noncommutativity ($\theta_e\neq 0$)  
\cite{AGBZ}. In Refs. \cite{SJ,Chaichian} it was argued that the
CPT theorem is also satisfied in these theories, based on the  
transformation properties of the different terms in the
classical action. What we see is that even if CPT is a  
symmetry at tree level, the full quantum theory is not necessarily invariant.

\subsection{Remarks on spin-statistics}

If we consider representations of SO(1,1)$\times$ SO(2) induced from string theory,
they are reductions of representations of SO(1,3). For them it is easy to extend
the above proof of the CPT theorem
\begin{eqnarray}
\langle\Omega|\Phi_{1}(x_{1})\ldots \Phi_{n}(x_{n})
|\Omega\rangle= (-1)^{J}i^{F}
\langle\Omega|\Phi_{n}(-x_{n})\ldots \Phi_{1}(-x_{1})
|\Omega\rangle
\label{cptGen}
\end{eqnarray}
where $F$ is the number of fermionic fields (which has to be even) and
$J$ is the number of undotted indices that results when the SO(1,1)$\times$SO(2)
representations are written in terms of representations of SO(1,3).

If instead one wants to consider representations of SO(1,1)$\times$SO(2) that do not follow
from restrictions of those of SO(1,3), the structure of the phases is more delicate. The 
kinetic term for the fields will generically be very different from the standard 
four-dimensional case, and although the proof of CPT can be adapted to this case,
the spin-statistics connection generically will fail.

For fully relativistic field theories, the spin-statistics theorem follow
from the phases appearing in the CPT theorem Eq. (\ref{cptGen}). In the noncommutative
case there are conceptual issues indicating that any spin-statistics theorem
will be more difficult to come by. In the standard case, and for massive theories,
the little group contains SO(3), a non-abelian group, which guarantees that the internal spin
$s$ of the physical states will be quantized according to $s\in \mathbb{Z}$ for bosons
and $s\in \mathbb{Z}+{1\over 2}$ for fermions. Thus, proving the spin-statistics connection
requires among other things to show that their interpolating fields satisfy microcausality with
commutators or anticommutators respectively\footnote{We are oversimplifying here the argument because 
there are subtle Klein factors that have to be taken into account in general (see \cite{Streater:vi}).}.
In the general noncommutative case these arguments fail, states with arbitrary helicity can be constructed, 
and it would not be surprising to find anyonic (or more exotic) behavior. For theories
whose field content descent from string theory via the Seiberg-Witten limit one expects
that the standard arguments for the spin-statistic theorem can be adapted to the noncommutative
case. The general case, however, remains to be elucidated.

\section{Additional issues}
\label{phenomenology}
\setcounter{equation}{0}

In theories with space-space noncommutativity, violations of CPT  
invariance can still appear for two reasons. The first one is the  
appearance of
tachyonic states in the spectrum which would spoil adapted weak local  
commutativity. The second thing that can go wrong
with the proof is that the Wightman  
functions do not define tempered distributions on the space
$\mathscr{S}[(\mathbb{R}^{1,3})^n]$. This is the case, for example, if
hard infrared singularities appear in the correlation functions induced  
by ultraviolet quadratic divergences via UV/IR mixing \cite{MvRS}.
If this happens, the analytic continuation of the Wightman functions  
into the (extended) tube might not be possible.

The two problems appear in the case of pure noncommutative  
QED with space-space noncommutativity ($\theta_e=0$, $\theta_{m}\equiv\theta$.),
where the one-loop corrected dispersion relation
presents a tachyonic instability at low energies \cite{QEDtachyon}
\begin{eqnarray}
\omega(\vec{p})^2=\vec{p}^{\,2}-{2g^2\over \pi^2 \,p\circ p}.
\end{eqnarray}
At the same time, the correlation functions are afflicted with  
infrared singularities derived from uncancelled quadratic ultraviolet
divergences. A regularization of these divergences by considering  
$\mathcal{N}=1$, U(1)$_{\star}$ gauge theory softly broken to
$\mathcal{N}=0$ does not eliminate the problem of the tachyonic  
states, since a negative, $\theta$-independent
squared mass for the photon is generated $m_{\rm photon}^2=-g^2  
M^2/(2\pi^2)$, with $M$ the soft breaking mass of the gaugino \cite{Carlson}.

To overcome these problems, we study U(1)$_\star$,  
$\mathcal{N}=4$ noncommutative gauge theory softly broken to
$\mathcal{N}=0$ by mass terms for the fermions and the scalars  
\cite{vdBij,KT}. This soft-breaking of supersymmetry introduces at most
logarithmic divergences in higher order amplitudes that do not destroy  
the tempered nature of the correlation functions \cite{soft}.

Before entering into the details of how this theory renders a  
non-tachyonic dispersion relation for the photon at low energies,
it is convenient to briefly discuss some aspects of the regularization of  
noncommutative gauge theories. In the usual case, 
the most popular regularization procedure is  
dimensional regularization, which straightforwardly extends to the
noncommutative case. However, in studying the Wilsonian low-energy  
effective action of noncommutative gauge theories, it would be convenient
to introduce a ``sharp'' cutoff in momentum space $\Lambda$. In this  
case the origin of the difficulties with
UV/IR mixing are relatively easy to understand: since the elementary  
objects are not particles but rigid dipoles of length  
$|\theta^{\mu\nu}p_{\nu}|$,
a ultraviolet cutoff in momenta $\Lambda$ induces an infrared cutoff  
$1/(\theta\Lambda)$, the inverse of the maximal dipole length.
Since at distances much larger than the length of the maximal dipole  
the elementary objects behave again like particles, the commutative
theory has to be recovered at momenta $|\vec{p}|\ll 1/(\theta\Lambda)$  
and at the same time Lorentz invariance is restored.

In the case of gauge theories, however, there is no obvious sharp momentum  
cutoff (apart from the lattice) compatible with gauge invariance and
the previous picture is not fully realized. Using dimensional  
regularizaton, for example, the whole region
$0\leq |\vec{p}|\leq 1/(\theta\Lambda)$
disappears. On the other hand, using a cutoff in the Schwinger  
parameter leads to violations of gauge invariance. This contrast with
the case, for example, of $\phi^{\star 3}$-theory in six-dimensions,  
where the picture outlined above is realized.

In order to avoid these problems we could use an ``intermediate''  
cutoff: we use dimensional regularization to regularize the
integrals along the directions orthogonal to the space-like vector  
$\tilde{p}^{\mu}\equiv \theta^{\mu\nu}p_{\nu}$ (with $p_{\mu}$ the
external momentum) while for the computation of the last integral  
along the direction of $\tilde{p}^{\mu}$ we use a Schwinger
cutoff. This procedure works for generic external momenta, i.e. $p\neq  
0$, $\tilde{p}\neq 0$.  Its advantage lies in the fact that
it exposes some of the features of the ``sharp'' cutoff explained  
above while preserving gauge invariance.

We proceed now to compute the one-loop effective action  
$\Gamma(\mathcal{A})_{\rm eff}$ for gauge field for a $\mathcal{N}=4$,
U(1) noncommutative gauge theory softly broken to $\mathcal{N}=0$ by  
mass terms for the fermions and scalars in the theory
\begin{eqnarray}
\Gamma(\mathcal{A})_{\rm eff}&=&-{1\over 2}\log\det\Delta_{\rm  
gauge}+\log\det\Delta_{\rm ghosts}+{n_{f}\over 2}\log\det\Delta_{f}
-{n_{s}\over 2}\log\det\Delta_{s} \nonumber \\
&\equiv& {1\over 2}\int {d^4 p\over (2\pi)^4}  
\mathcal{A}^{\mu}(p)\Pi_{\mu\nu}(p)\mathcal{A}^{\mu}(-p)+\mathcal{O}(\mathcal{A}^3),
\end{eqnarray}
where $n_{f}$ and $n_{s}$ denote respectively the number of Weyl  
fermions and real scalars in the theory.
For $\mathcal{N}=1$ noncommutative U(1) supersymmetric Yang-Mills we  
have $n_{f}=1$, $n_{s}=0$, whereas for
$\mathcal{N}=2$ one has $n_{f}=2$, $n_{s}=2$ and $n_{f}=4$, $n_{s}=6$  
for $\mathcal{N}=4$.

Following \cite{KT}, we use the background field method and work in  
Euclidean space. The photon self-energy can be
written as the sum of the planar and the nonplanar contribution as
\cite{planar}
\begin{eqnarray}
\Pi_{\mu\nu}(p)=\Pi_{\mu\nu}(p,\ell=0)-\Pi_{\mu\nu}(p,\ell=\tilde{p})
\label{pi}
\end{eqnarray}
with
\begin{eqnarray}
\Pi_{\mu\nu}(p,\ell)&=&2\sum_{j} \alpha_j\int {d^4 k\over (2\pi)^4}   
\left\{ d(j)\left[{(2k+p)_{\mu} (2k+p)_{\nu} \over
(k^2+m_{j}^2)[(k+p)^2+m_{j}^2]}-{2\delta_{\mu\nu}\over  
k^2+m_{j}^2}\right]\right. \nonumber \\
&+& \left. 4 C(j){p^2 \delta_{\mu\nu}-p_{\mu}p_{\nu} \over  
(k^2+m_{j}^2)[(k+p)^2+m_{j}^2]}\right\} e^{ik\cdot \ell}.
\label{effect}
\end{eqnarray}
The sum is over all states running in the loop (gauge fields, ghosts,  
fermions and scalars). The constant $\alpha(j)$,
$d(j)$ and $C(j)$, as well as the masses for the different fields,  
are given in the following table:
\begin{eqnarray*}
\begin{array}{|c|c|c|c|c|}
\hline
   &     &     &     &    \\
j  &  {\rm ghost}  &  \mbox{real scalar}  &  \mbox{Weyl fermion}  &   
\mbox{gauge field} \\
   &     &     &     &    \\
\hline
   &     &     &     &    \\
\alpha_j &  1  & -{1\over 2}  &  {1\over 2}   &  -{1\over 2}  \\
       &    &    &   &   \\
d(j)  & 1   & 1  & 2  &  4    \\
       &     &     &    &     \\
C(j)  & 0  &  0  & {1\over 2} & 2 \\
  &     &     &     &    \\
m_{j}  & 0  &  M_{s} & M_{f} & 0 \\
  &     &     &     &    \\
\hline
\end{array}
\end{eqnarray*}
In theories with unbroken (or softly broken) supersymmetry the condition ${\rm Str\,}\mathbf{1}=0$
translates into 
\begin{eqnarray}
\sum_{j}\alpha_{j}d(j)=n_{f}-{n_{s}\over 2}-1=0
\label{str1}
\end{eqnarray}
whereas in theories with $\mathcal{N}=4$ (or $\mathcal{N}=2$ with two  
hypermultiplets) we have the additional identity
\begin{eqnarray}
\sum_{j}\alpha_{j}C(j)\equiv {n_{f}\over 4}-1=0.
\label{n=4}
\end{eqnarray}
For supersymmetric theories (\ref{str1}) guarantees the vanishing of the terms proportional 
to $\tilde{p}_{\mu}\tilde{p}_{\nu}$ in the vacuum polarization, and the dispersion relation of 
the photon is not modified. In the particular case of noncommutative gauge theories with
$\mathcal{N}=4$ unbroken supersymmetry the identity (\ref{n=4}) further implies that $\Pi_{\mu\nu}(p)=0$.
If soft breaking masses are
included these cancellations are not complete but, as we will see below,  
they are enough to tame the problems arising from
UV/IR mixing.

In evaluating (\ref{effect}) we come back to the issue of  
regularization in more detail. There are various ways to regulate
the planar diagram contribution $(\ell=0)$, but few for the non-planar  
part ($\ell=\tilde{p}$) preserving  gauge invariance.
Physically, when working with the low-energy Wilsonian effective  
action it is convenient to introduce some kind of
sharp cutoff $\Lambda$ which eliminates the physics at scales  
$E>\Lambda$. In theories without gauge symmetry, and at one
loop, this can be achieved by exponentiation of the propagators using  
Schwinger parameters and modifying their integration
measure. For instance, for two propagators the procedure amounts to the prescription
\begin{eqnarray}
{1\over  
(p_{1}^2+m_{1}^2)(p_{2}^2+m_{2}^2)}&=&
\int_{0}^{\infty}d\alpha_{1}d\alpha_{2}\exp\left[-\alpha_{1}(p_{1}^2+m_{1}^2)
-\alpha_{1}(p_{2}^2+m_{2}^2)\right]  \\
&\hspace*{-0.9cm}\longrightarrow\hspace*{0.3cm} & \hspace*{-0.5cm}\int_{0}^{\infty}d\alpha_{1}d\alpha_{2}  
\exp\left[-\alpha_{1}(p_{1}^2+m_{1}^2)
-\alpha_{1}(p_{2}^2+m_{2}^2)-{1\over  
4\Lambda^2(\alpha_{1}+\alpha_{2})}\right].
\nonumber
\end{eqnarray}
Proceeding along these lines, Eqs. (\ref{pi}) and (\ref{effect})  
become after some computation
\begin{eqnarray}
\Pi_{\mu\nu}(p)&=& {1 \over  
4\pi^2}\left(p^2\delta_{\mu\nu}-p_{\mu}p_{\nu}\right) \nonumber \\
&  & \,\,\times \,\,\sum_{j}\alpha_{j}\int_{0}^{1}dx\,
\left[4C(j)-(1-2x)^2 d(j)\right]
\left[K_{0}\left({\sqrt{\Delta_{j}}\over \Lambda}\right)
-K_{0}\left({\sqrt{\Delta_{j}}\over \Lambda_{\rm eff}}\right)\right]  
\nonumber\\
&+& {1\over (4\pi)^2}\tilde{p}_{\mu}\tilde{p}_{\nu}\,\Lambda_{\rm  
eff}^2\sum_{j}\alpha_{j}d(j)
\int_{0}^{1}dx\,\Delta_{j}K_{2}\left({\sqrt{\Delta_{j}}\over  
\Lambda_{\rm eff}}\right) \nonumber \\
&+& \delta_{\mu\nu}\left[\mbox{ gauge non-invariant term }\right],
\label{answer}
\end{eqnarray}
where $\Delta_{j}=m_{j}^2+x(1-x)p^2$ and the effective cutoff is given  
by \cite{MvRS}
\begin{eqnarray}
{1\over \Lambda_{\rm eff}^{2}}={1\over \Lambda^2}+\tilde{p}^2
\end{eqnarray}
and we have expressed the integrals over Schwinger parameters in terms  
of modified Bessel functions of the second kind
\begin{eqnarray}
K_{\nu}(\alpha z)={\alpha^\nu\over 2}\int_{0}^{\infty}{dt\over  
t^{\nu+1}} e^{-{z\over 2}\left(t+{\alpha^2\over t}\right)}.
\end{eqnarray}
Ignoring momentarily the gauge non-invariant term, the answer  
(\ref{answer}) is satisfactory in several ways.
For fixed $\Lambda$, in the momentum region  
$|\vec{p}|<1/(\Lambda\theta)$ we recover Lorentz invariance and the  
standard dispersion
relation for the photon. Also there is no ambiguity in taking the  
limit $\tilde{p}\rightarrow 0$. The big drawback, of course,
is the lack of gauge invariance, due to the term proportional to  
$\delta_{\mu\nu}$.
This term can only be subtracted by local  
counterterms for the planar diagram.

Had we used dimensional regularization, the answer would be given by  
the leading term of (\ref{answer}) in the limit $\Lambda\rightarrow
\infty$, at fixed $\tilde{p}$,
\begin{eqnarray}
K_{0}\left({\sqrt{\Delta_{j}}\over \Lambda}\right) \longrightarrow   
-{1\over 2}\log \left({\Delta_{j}\over 4\pi\mu^2}\right);
\hspace*{1cm}
\Lambda_{\rm eff} \longrightarrow  {1\over |\tilde{p}|}
\end{eqnarray}
and with the gauge non-invariant term absent (cf. \cite{MRR}). In this  
limit the region $0<|\vec{p}|<1/(\Lambda\theta)$ disappears.

It is possible to imagine an intermediate regulator where the loop  
integrals are split into a three-dimensional part, which is
evaluated using dimensional regularization in $3-\epsilon$ dimensions,  
and a fourth integral which is regularized using a Schwinger
cutoff. In order to implement this regularization we consider for  
generic $p$ (i.e. $p\neq 0$, $\tilde{p}\neq 0$) an orthonormal
frame ${\rm e}_{\hat{a}\mu}$ ($\hat{a}=1,\ldots,4$)
\begin{eqnarray}
{\rm e}_{\hat{1}\mu}={\tilde{p}_{\mu}\over |\tilde{p}|}, \hspace*{0.5cm} 
{\rm e}_{\hat{2}\mu}={p_{\mu}\over |p|}
\end{eqnarray}
and ${\rm e}_{\hat{3}\mu}$, ${\rm e}_{\hat{4}\mu}$ chosen so that  
${\rm e}_{\hat{a}\mu}\cdot{\rm e}_{\hat{b}\nu}=
\delta_{\hat{a}\hat{b}}$.
In this case Eq. (\ref{effect}) can be split into an integral over  
$k_{\hat{1}}$, which is regulated using a Schwinger cutoff
(physically, this corresponds
to regulating the length of the dipoles running in the loop),
and the remaining three-dimensional integral, which can be dealt with  
using dimensional regularization. 

Implementing this ``mixed'' regulator we find a result for the  
polarization tensor which is not identical to (\ref{answer}),
but in which the gauge non-invariant term is absent and the effective  
cutoff $\Lambda_{\rm eff}$ appearing in the third line of
this equation is replaced by $(2\tilde{p}^2\Lambda_{\rm  
eff}^2-1)/\tilde{p}^2$.
If we take the limit $\Lambda\rightarrow \infty$ with $\tilde{p}$ fixed,
we recover the result of dimensional  
regularization, again with the identification $\Lambda^2\rightarrow  
4\pi\mu^2$.
The main problem with this procedure is that it is not clear whether  
it can be systematically extended to higher loops.

\subsection{Softly broken $\mathcal{N}=4$, U(1)$_{\star}$ gauge theory}

Let us now focus on the case of $\mathcal{N}=4$, noncommutative U(1)  
gauge theory. Using dimensional regularization,
the polarization tensor can be written as
\begin{eqnarray}
\Pi_{\mu\nu}(p)=\Pi_{1}(p)(p^2\delta_{\mu\nu}-p_{\mu}p_{\nu})+\Pi_{2}(p){\tilde{p}_{\mu}\tilde{p}_{\nu}\over  
\tilde{p}^2},
\label{dr}
\end{eqnarray}
where
\begin{eqnarray}
\Pi_{1}(p)&=&{1\over  
4\pi^2}\int_{0}^{1}dx\left\{\left[4-(1-2x)^2\right]\left[{1\over  
2}\log\left({\Delta_{v}\over 4\pi\mu^2}\right)
+K_{0}(\sqrt{\Delta_{v}}|\tilde{k}|)\right] \right.\nonumber \\
&-& \left[1-(1-2x)^2\right]\sum_{f}\left[{1\over  
2}\log\left({\Delta_{f}\over 4\pi\mu^2}\right)
+K_{0}(\sqrt{\Delta_{f}}|\tilde{k}|)\right] \nonumber \\
&-& \left.{1\over 2}(1-2x)^2\sum_{s}\left[{1\over  
2}\log\left({\Delta_{s}\over 4\pi\mu^2}\right)
+K_{0}(\sqrt{\Delta_{v}}|\tilde{k}|)\right]\right\}
\label{pi1}
\end{eqnarray}
and 
\begin{eqnarray}
\Pi_{2}(p)&=& -{1\over  
\pi^2}\int_{0}^{1}dx\,\left[\Delta_{v}K_{2}(\sqrt{\Delta_{v}}|\tilde{p}|)-
\sum_{f}\Delta_{f}K_{2}(\sqrt{\Delta_{f}}|\tilde{p}|) \right.\nonumber  
\\
&+& \left.{1\over  
2}\sum_{s}\Delta_{s}K_{2}(\sqrt{\Delta_{s}}|\tilde{p}|)\right]
\label{pi2}
\end{eqnarray}
The subindices $v$, $f$ and $s$ indicate respectively the  
contributions of the vector-ghost system, fermions and scalars, and
\begin{eqnarray}
\Delta_{v}=x(1-x)p^2, \hspace*{1cm} \Delta_{f}=m_{f}^2+x(1-x)p^2,  
\hspace*{1cm}
\Delta_{s}=m_{s}^2+x(1-x)p^2.
\end{eqnarray}

The Wilsonian effective coupling constant at momentum $p$ is  
determined by $\Pi_{1}(p)$, namely
\begin{eqnarray}
{1\over g(p)^2}-{1\over g_{0}^2}=\Pi_{1}(p).
\end{eqnarray}
For values of the momenta larger that the noncommutative scale, $p\gg  
1/\sqrt{\theta}$, the Bessel functions decay exponentially
and only the logarithms in (\ref{pi1}) contribute, reproducing the  
standard $\beta$-function of the theory
\begin{eqnarray}
\Pi_{1}(p)\simeq {1\over 8\pi^2}\left({11\over 3}-{2\over  
3}n_{f}-{1\over 6}n_{s}\right)\log\left({p^2\over 4\pi\mu^2}\right),
\hspace*{1cm} |p|\gg {1\over \sqrt{\theta}}.
\label{uv}
\end{eqnarray}
If we have a
field content corresponding to softly broken $\mathcal{N}=4$ U(1)  
theory,  a $\mathcal{N}=1$ vector multiplet and three chiral
multiplets in the adjoint representation, the $\beta$-function  
vanishes. In fact the theory is believed to be ultraviolet finite \cite{JJ}.

Before considering the modifications that arise from the introduction  
of the soft breaking masses, we consider the supersymmetric
case where $M_{f}=M_{s}=0$. In this case there is an interesting  
phenomenon associated with UV/IR mixing. If we consider the region
of small  momenta $p\ll 1/\sqrt{\theta}$ the Bessel function can be  
approximated by its leading logarithm behavior and therefore
we have
\begin{eqnarray}
-{1\over 2}\log\left({\Delta\over  
4\pi\mu^2}\right)-K_{0}(\sqrt{\Delta}|\tilde{p}|) \simeq {1\over  
2}\log(4\pi\mu^2 \tilde{p}^2)
\end{eqnarray}
Therefore, in this limit we find for $\Pi_{1}(p)$
\begin{eqnarray}
\Pi_{1}(p)\simeq {1\over 8\pi^2}\left(-{11\over 3}+{2\over  
3}n_{f}+{1\over 6}n_{s}\right)\log(4\pi\mu^2\tilde{p}^2),
\hspace*{1cm} |p|\ll {1\over \sqrt{\theta}}
\label{ir}
\end{eqnarray}
Comparing Eqs. (\ref{uv}) with (\ref{ir}) we find that the running of  
the coupling constant in the infrared is completely similar
to the one in the ultraviolet, except for a change in the sign. The  
different sign indicates that, for theories with a negative
$\beta$-function, the theory becomes weakly coupled again at low  
energies. This kind of duality in momenta, $p\leftrightarrow  
1/\tilde{p}$,
is a reflection of the mixing of high and low energy scales  
characteristic of noncommutative field theories.

We next proceed to include the soft breaking masses for the fermions  
and the scalars. In particular, we will consider the $\mathcal{N}=4$
theory in the region of momenta $|p|\ll m_{j}<1/\sqrt{\theta}$. In this  
case, $\Pi_{1}(p)$ can be written as
\begin{eqnarray}
\Pi_{1}(p)={1\over 8\pi^2}\left(-{11\over 3}\right)\log\left({  
M_{\rm eff}^2\tilde{p}^2}\right) + \mbox{exponentially vanishing terms}
\end{eqnarray}
where the effective mass is given by
\begin{eqnarray}
M_{\rm eff}=\left(\prod_{f=1}^{4}M_{f}^2\prod_{s=1}^{6}M_{s}^{1\over  
2}\right)^{1\over 11},
\end{eqnarray}
again leading to a weakly coupled theory in the infrared.  Hence from
this point of view the Wilsonian effective action is well defined.  At
low-energies we recover asymptotically an effective theory of a photon
whose coupling constant vanishes in the infrared.  
These conclusions hold as long as there are no tachyons in the spectrum,
which among other things destroy the possibility of having a weak version
of local commutativity needed in the proof of the CPT theorem, as already discussed in 
Section \ref{cpt}.

As we mentioned, pure $\mathcal{N}=1$, U(1)$_{\star}$   with a mass term for
the photino leads to a low-energy tachyon pole for the photon
\cite{Carlson}.  In our case, since we can also play with the
scalar masses, it is possible to choose values of the soft breaking masses
for the fermions and the scalar in such a way that we avoid this
problem.  To understand the photon dispersion relation we
need to compute the full propagator following from (\ref{dr}).

In order to find the poles of the full propagator we should resum the two-point
1PI diagrams. The structure of (\ref{dr}) suggest the introduction of the non-orthogonal
projectors
\begin{eqnarray}
P_{\mu\nu}&=& \left(\delta_{\mu\nu}-{p_{\mu}p_{\nu}\over p^2}\right), \nonumber \\
Q_{\mu\nu}&=& {\tilde{p}_{\mu}\tilde{p}_{\nu}\over \tilde{p}^2}
\end{eqnarray}
which satisfy (in matrix notation)
\begin{eqnarray}
\mathbf{P}^{n}=\mathbf{P}, \hspace*{1cm} 
\mathbf{Q}^{n}=\mathbf{Q}, \hspace*{1cm} 
\mathbf{PQ}=\mathbf{QP}=\mathbf{Q}
\label{pq}
\end{eqnarray}
so the polarization tensor in Eq. (\ref{dr}) can be written
\begin{eqnarray}
\mathbf{\Pi}(p)= p^2\Pi_{1}(p)\mathbf{P}+\Pi_{2}(p)\mathbf{Q}.
\end{eqnarray}
The full propagator can then be written as
\begin{eqnarray}
G_{\mu\nu}(p)={ig_{0}^2\over p^2}\left[\mathbf{1}+{-g_{0}^2\over p^2}\mathbf{\Pi}(p)
+\left({-g_{0}^2\over p^2}\right)^2\mathbf{\Pi}(p)^2+\ldots\right]_{\mu\nu}.
\end{eqnarray}
In order to resum the series, we use Eq. (\ref{pq}) to write
\begin{eqnarray}
\mathbf{\Pi}(p)^n=\left[p^2\Pi_{1}(p)\right]^n \mathbf{P}+\left\{
\left[p^2\Pi_{1}(p)+\Pi_{2}(p)\right]^n-\left[p^2\Pi_{1}(p)\right]^n\right\}\mathbf{Q}
\end{eqnarray}
so the full propagator is given by
\begin{eqnarray}
G_{\mu\nu}(p)&=&{ig_{0}^2 p_{\mu}p_{\nu}\over p^2}
+{ig_{0}^2\over p^2[1+g_{0}^2\Pi_{1}(p)]}\left(\delta_{\mu\nu}-{p_{\mu}p_{\nu}\over p^2}\right) 
\nonumber \\
&+&\left\{{ig_{0}^2\over p^2\left[1+g_{0}^2\Pi_{1}(p)\right]+g_{0}^2\Pi_{2}(p)}
-{ig_{0}^2\over p^2[1+g_{0}^2\Pi_{1}(p)]}\right\}{\tilde{p}_{\mu}\tilde{p}_{\nu}
\over \tilde{p}^2}.
\label{fullprop}
\end{eqnarray}

Therefore, from Eq. (\ref{fullprop}) we find that the pole conditions are
\begin{eqnarray}
p^2 \left[{1\over g_0^2}+\Pi_1(p)\right] &=& 0 
\label{poles1}
\end{eqnarray}
and
\begin{eqnarray}
p^2 \left[{1\over g_0^2}+\Pi_1(p)\right] + \Pi_2(p)&=&0.
\label{poles2}
\end{eqnarray}
To obtain the dispersion relations for physical particles
we have to rotate to Minkowski space by taking $p^2\rightarrow -p^2$ and
$\tilde{p}^2\rightarrow p\circ p$.  Equation (\ref{poles1})
implies that one of the polarizations of the photon
corresponds to a massless degree of freedom.  On the other hand, Eq. (\ref{poles2})
is associated to a photon whose polarization is  proportional to 
${\tilde p}^{\mu}$.
To determine the pole in this case we need the low energy behavior of
$\Pi_2(p)$. From (\ref{pi2}), and using the small argument expansion
of $K_2(z)$, we obtain after simple manipulations
\begin{eqnarray}
\Pi_2(p) &=&  {1\over \pi^2}\left(-1+ n_f -{n_s\over 2}\right)
\left({2\over {\tilde p}^2}-{p^2\over 12}\right)
-
{1\over 2\pi^2}\left(\sum_f M_f^2-{1\over 2}\sum_s M_s^2\right)
+\ldots
\end{eqnarray}
In $\mathcal{N}=4$ noncommutative super Yang-Mills,  we have
$-1+n_f+{n_s\over 2}=0$.  Hence to avoid tachyons we
need to satisfy 
\begin{eqnarray}
\sum_f M_f^2-{1\over 2}\sum_s M_s^2 \leq 0
\label{ineq}
\end{eqnarray}
which implies ${\rm Str\,}M^2\geq 0$.
This condition guarantees, to this order in perturbation theory, that the spectrum is 
free of tachyons. Although when the equality in (\ref{ineq}) is satisfied we recover
formally a massless photon at zero momentum, the function $\Pi_{2}(p)$ is negative in a 
neighborhood of $\tilde{p}=0$. Therefore, in order to avoid problems the soft-breaking
masses has to be tuned so the photon has a positive mass squared at zero momentum.

\begin{figure}[ht]
\centerline{ \epsfxsize=4.0truein \epsfbox{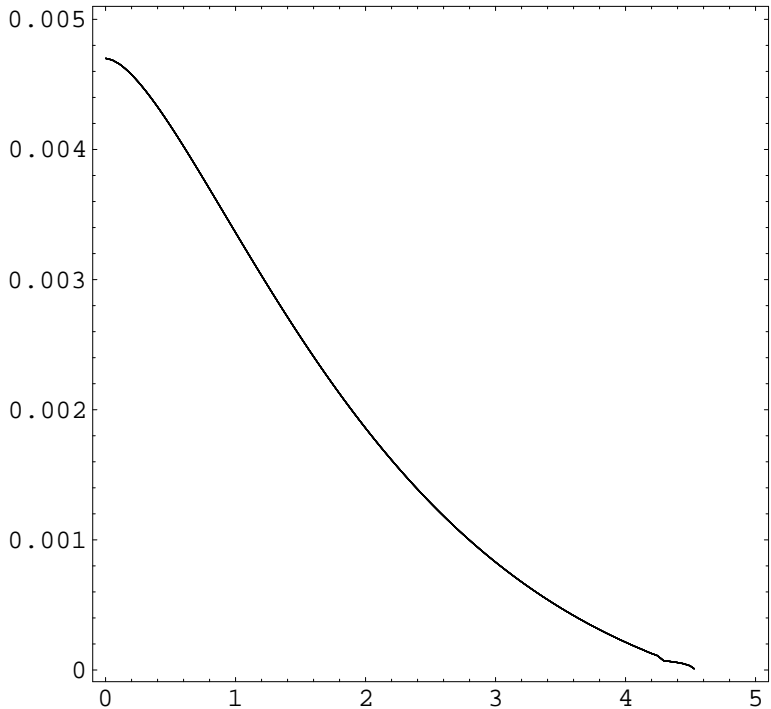}}
\caption{Dispersion relation $p^2$ versus $t$.}
\end{figure}
In Fig. 5 we have plotted the dispersion relation by solving (\ref{poles2}) as an
implicit equation for $p^2\equiv s-t$ and $t\equiv p\circ p/\theta^2$, when the strict inequality in
(\ref{ineq}) is satisfied. 
The positive intercept in the curve implies that the photons with this polarization
become massive, something that phenomenologically is a disaster.
The current bound on the photon mass is
$m_{\gamma} < 2\times 10^{-16}$ eV \cite{PDG}, hence to satisfy
it one would have to do a rather non-trivial fine tuning in the
expansion of $\Pi_2(p)$. However, even if this is
achieved, the dispersion relation is likely to produce birefringence,
i.e. the speed of light would be different for polarizations 
along the commutative and noncommutative directions. This effect
would come from the terms in $\Pi_2(p)$ quadratic in $p_{\mu}$ and
${\tilde p}_{\mu}$.  Again in this case the bounds are very restrictive.
Another very strict bound can also be extracted by looking at the departures from the black body
radiation that a massive polarization for the photon produces.

It is clear that the phenomenological prospects for this theory
are rather slim. The main purpose to study it was to show that it
is possible to construct a noncommutative field theory satisfying
the adapted axioms introduced in Section \ref{adapted}.  At least in perturbation,
theory the correlation functions calculated are tempered distributions,
the theory is well defined both in the ultraviolet and the infrared,
and the renormalization procedure and the computation of the Wilsonian
effective action will not be afflicted by hard infrared divergences
that would lead to Wightman functions with non-tempered singularities.
This is for instance the reason why massless scalar fields in two-dimensions
do not exist \cite{wightcargese}.

Before closing this section, we make some remarks about the resulting dispersion relations using 
other regularizations procedures mentioned above involving a sharp cutoff. 
Modulo the problems already pointed out, in this
case we find that for the photons with polarizations along $\tilde{p}^{\mu}$ the usual dispersion
relation $p^2=0$ is recovered for both low and high noncommutative momenta with respect 
to the noncommutative scale $1/\sqrt{\theta}$. Around this noncommutative scale, however, 
one finds a region where the group velocity of waves packets becomes superluminal.
This situation is highly reminiscent of the situation described in \cite{plasma} for noncommutative
field theories at finite temperature. This is however not so surprising if one keeps in
mind that in thermal noncommutative field theory the temperature plays the role of a ``sharp'' cutoff. 
Indeed, at fixed $T$ a noncommutative U(1)$_{\star}$ gauge theory 
has a smooth infrared (and commutative) limit \cite{AVM} due to the fact that the Boltzmann
factors in the temperature-dependent part of the loop integrals cut 
off any physics above the scale $T$, thus regularizing the UV/IR mixing of
noncommutative gauge theories.

Although we have concentrated only on the one-loop vacuum polarization for the
photon, we expect the same conclusions to apply to higher orders, and also to
higher point functions.  The theory considered is believed to be finite in the
ultraviolet, and once the possibility of tachyon poles is allayed, the correlation
functions for the theory should be described in terms of tempered distributions,
and the low-energy Wilsonian effective action should also be well defined.
Nevertheless, the constraint between the soft-breaking masses (\ref{ineq}) is very 
likely to be modified by the inclusion of higher loop corrections.

\section{Concluding remarks}
\label{conclusions}
\setcounter{equation}{0}

In this paper we have studied some general properties of noncommutative
quantum field theories. An axiomatic formulation can be 
achieved by modifying the standard Wightman axioms
using as guiding principles i) the breaking of Lorentz symmetry down to
the subgroup O(1,1)$\times$SO(2), and ii) the relaxation of local commutativity
to make it compatible with the causal structure of the theory, given by the light-wedge
associated with the O(1,1) factor of the kinematical symmetry group. 
 
These axioms are enough to demonstrate the CPT theorem. Indeed, this theorem holds
for these theories for reasons not very different from the ones behind the 
CPT theorem for commutative theories. The transformation PT is in the connected 
component of the identity in the group that result from the complexification of
the O(1,1) subgroup of the symmetry group. This, together with the tempered character of
the distributions defined by the Wightman functions, allows to use results like the
``edge of the wedge" theorem to show that the Wightman functions and their
CPT-transforms coincide. 
 
The main source of difficulties in formulating noncommutative 
field theories which satisfy the adapted axioms is UV/IR mixing. The existence 
of hard infrared singularities in the non-planar sector of the theory, induced by 
uncancelled quadratic ultraviolet divergences, can result in two kinds of problems: 
they can destroy the tempered nature of the Wightman functions and/or they can 
introduce tachyonic states in the spectrum, so the modified postulate of local commutativity
is not preserved.
 
In this sense, we have also shown how it is possible to construct noncommutative 
field theories which, at least in perturbation theory, are compatible with the adapted axioms. 
Noncommutative QED is an example of a theory which does not satisfy the axioms 
due to the emergence of tachyons due to UV/IR mixing. This mixing of scales can be tamed
by completing noncommutative QED to $\mathcal{N}=4$ U(1)$_{\star}$ super Yang-Mills
and breaking supersymmetry softly by introducing masses for the scalar and fermion fields.
This eliminate the quadratic divergences and, if the soft-breaking masses are tuned, 
remove tachyonic instabilities at low momentum. Similar arguments will apply to higher
orders in perturbation theory and higher point functions since the underlying theory
is finite ${\cal N}=4$ noncommutative QED.  We have also shown that in this context
the U(1) theory obtained is a phenomenological disaster.  There are other approaches
to phenomenology in noncommutative field theories (see for instance \cite{trampetic}),
where one first expands the action up to a certain order in $\theta$ and then quantizes
the theory obtained.  We know, however, that generally the two procedures do not necessarily
generate the same results due to the UV/IR mixing.

\section*{Acknowledgments}

We would like to thank Jos\'e L.~F.~Barb\'on, Jos\'e M.  
Gracia-Bond\'{\i}a, Kerstin E. Kunze, Dieter L\"ust, Raymond Stora and 
Julius Wess for useful discussions. We would like to thank specially 
Masud Chaichian for discussions about noncommutative theories and the CPT theorem
which prompted our interest in the subject. Both authors  
wish to thank the Humboldt  
Universit\"at zu Berlin, and in particular Dieter L\"ust, for
hospitality during the completion of this work. M.A.V.-M. acknowledges the 
support of Spanish Science Ministry Grants AEN99-0315 and FPA2002-02037.

\end{document}